\documentclass[preprintnumbers,twocolumn,groupedaddress,nofootinbib,aps]{revtex4}
\usepackage{float}
\usepackage{amsmath,amssymb}
\usepackage{graphicx}
\usepackage{color}
\usepackage{hyperref}
\usepackage{url}
\usepackage{slashed}
\usepackage{subfigure}
\usepackage{slashed,bbm}
\usepackage{graphics,psfrag,epsfig}
\usepackage{dsfont}
\usepackage{enumerate}
\usepackage{textcomp}
\usepackage{bm}
\usepackage{lmodern}
\usepackage{lmodern}
\usepackage[latin9]{inputenc}
\usepackage{bm}
\usepackage{amsmath}
\usepackage{amssymb}
\usepackage{graphicx}
\usepackage{esint}
\usepackage{verbatim}

\newcommand{\RNum}[1]{\uppercase\expandafter{\romannumeral #1\relax}}

\newcommand{\beq}{\begin{equation}}
\newcommand{\eeq}{\end{equation}}
\newcommand{\bea}{\begin{eqnarray}}
\newcommand{\eea}{\end{eqnarray}}
\newcommand{\be}{\begin{equation}}
\newcommand{\ee}{\end{equation}}

\newcommand{\Ut}[1]{\tilde{U}}
\newcommand{\Utt}[1]{\tilde{\tilde{U}}}

 \definecolor{BLACK}{gray}{0}
 \definecolor{WHITE}{gray}{1}
 \definecolor{RED}{rgb}{1,0,0}
 \definecolor{GREEN}{rgb}{0,1,0}
 \definecolor{BLUE}{rgb}{0,0,1}
 \definecolor{CYAN}{cmyk}{1,0,0,0}
 \definecolor{MAGENTA}{cmyk}{0,1,0,0}
 \definecolor{YELLOW}{cmyk}{0,0,1,0}

\makeatother

\begin{document}

\title{Orbital order in FeSe -- the case for vertex renormalization}

\author{Rui-Qi Xing$^{1}$, Laura Classen$^{2}$ and Andrey V. Chubukov$^{1}$}
\affiliation{$^{1}$ School of Physics and Astronomy, University of Minnesota, Minneapolis, MN 55455, USA \\ $^{2}$ Condensed Matter Physics and Materials Science Department, Brookhaven National Laboratory, Upton, NY 11973-5000, USA}

\begin{abstract}
We study the structure of the d-wave orbital order in FeSe in light of recent STM and ARPES data, which detect the shapes of hole and electron pockets in the nematic phase. The geometry of the pockets  indicates that the sign of the orbital order $\Gamma =
\langle d^{\dagger}_{xz} d_{xz}- d^{\dagger}_{yz} d_{yz}\rangle$ is different between hole and electron pockets $(\Gamma_h$ and $\Gamma_e$).  We argue that this sign change cannot be reproduced if one
 solves for the orbital order within mean-field approximation, as the mean-field analysis yields either no orbital order, or order with the same sign of $\Gamma_e$ and $\Gamma_h$.   We argue that another solution with the opposite signs of $\Gamma_e$ and $\Gamma_h$  emerges if we include the  renormalizations of the vertices in $d-$wave orbital channel. We show that the ratio $|\Gamma_e/\Gamma_h|$ is of order one, independent on the strength of the interaction. We also compute the  temperature variation of the energy of $d_{xz}$ and $d_{yz}$ orbitals at the center of electron pockets and compare the results with ARPES data.
  \end{abstract}

\maketitle

{\it \bf Introduction}~~~Orbital degrees of freedom turned out to play an important role for iron-based superconductors (FeSC).
 Studies of SDW magnetism and superconductivity in these materials found that the orbital composition of the states near the Fermi surface (FS) affects the structure of the fermionic spectrum in the spin-density-wave (SDW) phase~\cite{vishw} and the anisotropy of the superconducting gap
  \cite{scalapino,arakawa2011,ckf}.
  Another example where different orbitals come into play  is the tetragonal-to-orthorhombic phase transition observed in many FeSCs at $T = T_s$. Below $T_s$, the system spontaneously breaks $C_4$ lattice rotational symmetry down to $C_2$. This is similar to what happens in nematic liquid crystals, and, by analogy, the state below $T_s$ is called nematic.  Below $T_s$ the occupation of $d_{xz}$ and $d_{yz}$ orbitals becomes unequal, i.e., the system develops an orbital order $\Gamma (|{\bf k}|) \propto \int d\theta_k  (n_{xz} ({\bf k}) - n_{yz} ({\bf k}))$,
  where $n_i$ is the density of orbital $i$
  and the integration is over the directions of ${\bf k}$ for a given $|{\bf k}|$.
    Above  $T_s$, $\Gamma_k$ vanishes by $C_4$ symmetry, but once $C_4$ symmetry is broken, by one reason or the other~\cite{review_ch_fern}, $\Gamma (|{\bf k}|)$ becomes finite.

  \begin{figure}[t]
	\centering{}
	\includegraphics[width=0.45\columnwidth]{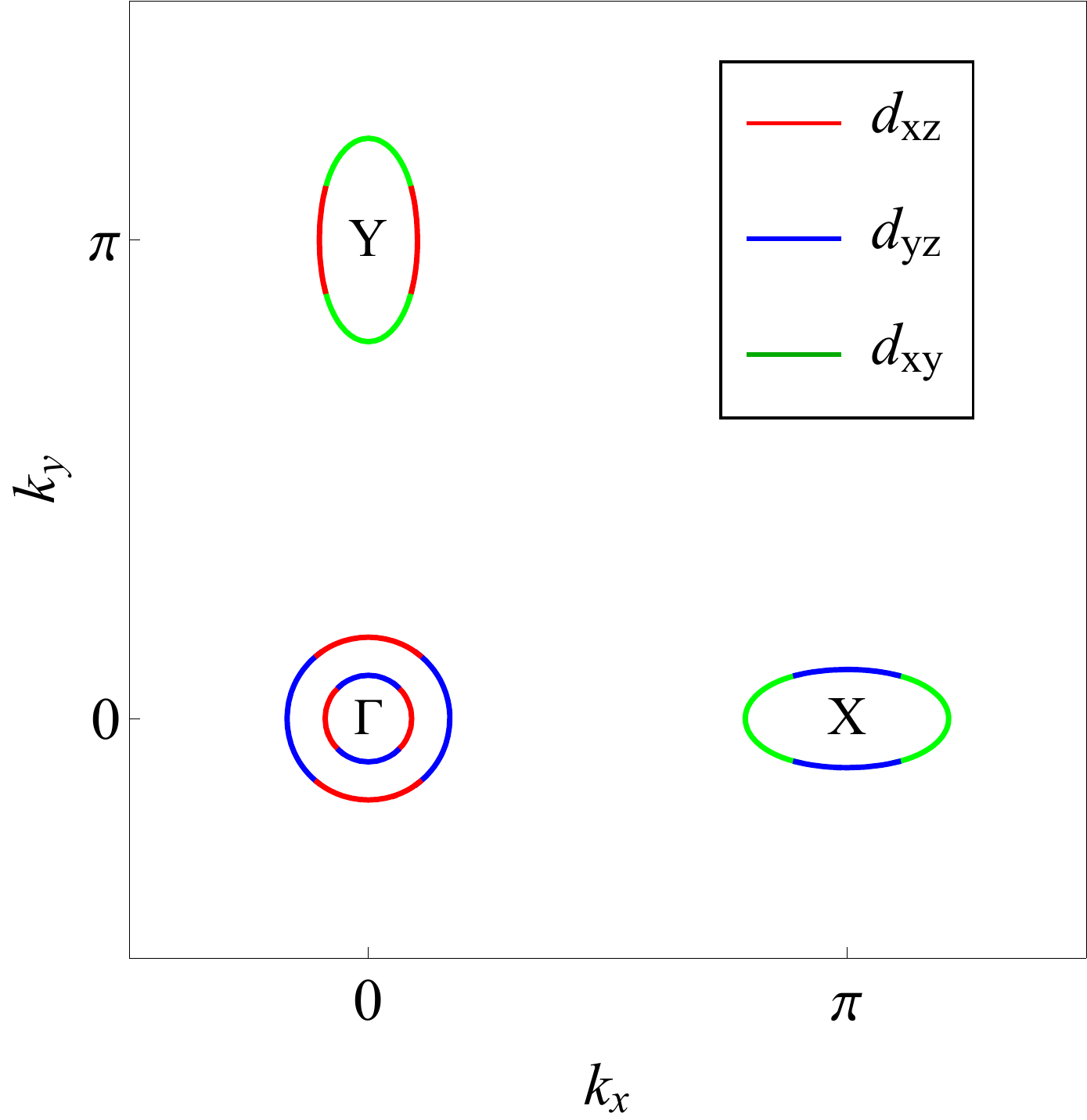} \quad
	\includegraphics[width=0.45\columnwidth]{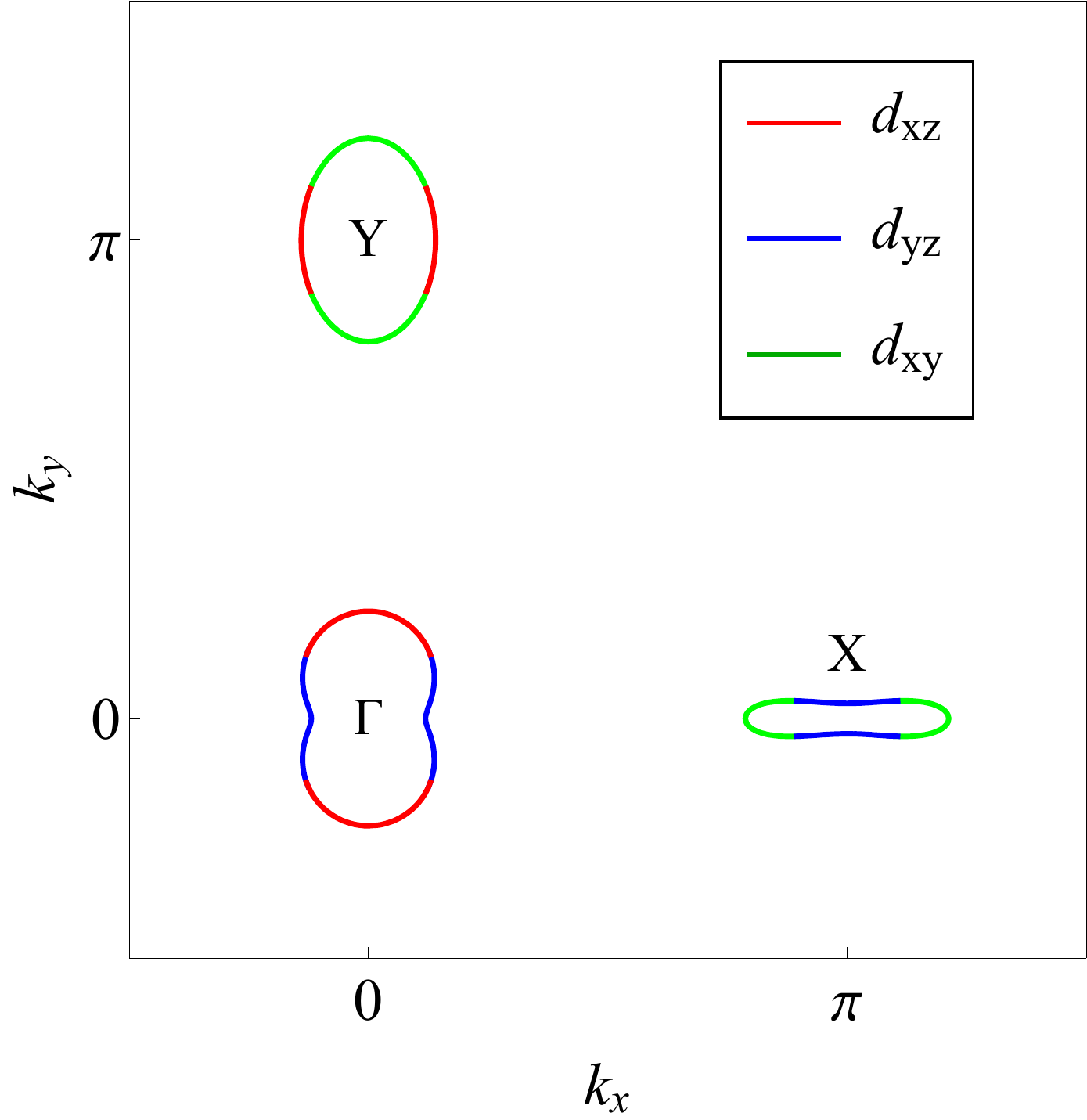}
	\protect\caption{The Fermi surfaces in the 1Fe BZ with the leading orbital content encoded in color.   The six $\psi$ fields, marked in the figure, are introduced in Table 1.  Left panel -- the FSs in the tetragonal phase,  right panel -- the FSs in the nematic phase.  The smaller hole pocket shrinks by orbital order and completely disappers once one include spin-orbit coupling.  In FeSe, the size of the larger hole pocket depends on $k_z$ and is the largest at $k_z =\pi$ (Ref. \cite{borisenko_latest}). The parameters of the quadratic Hamiltonian used to obtain the Fermi surface are from Ref.\cite{kang_latest}. The pockets are homogeneously enlarged to provide a better view.}
		\label{fig:FS_Sketch}
\end{figure}

 In most FeSCs  the range of nematic order is quite narrow as the system develops a stripe magnetic order almost immediately after the nematic order sets in.  However, in FeSe (and in doped FeSe$_{1-x}$S$_x$) the regions of nematic and magnetic order are well separated in $x$ (\cite{Shibauchi_1,recent})

  In  pure FeSe, the nematic order  sets in at $T_s\approx 85K$, and magnetic order does not develop down to $T=0$.  This opens up an opportunity to extract the information about the structure of  $\Gamma$  from the analysis of the  feedback effects on the  electronic structure.
  The magnitude of $\Gamma$, extracted from ARPES, is $10-20 meV$, much smaller than the fermionic bandwidth (see Refs \cite{recent,Fedorov,Coldea_AnnPhy,davis,arpes_strained,watson,shibauchi,Zhou,comm2}
   and the discussion below).  In this case, the most relevant feedback from $\Gamma$ on the electrons is for momentum components in the XY plane  near $k_x=k_y =0$ and $k_x = k_y = \pi$, where hole and electron pockets are located in the 2Fe Brillouin zone (2FeBZ).  The pockets in FeSe are quite small, and  $\Gamma (|{\bf k}|)$ near these pockets is well approximated by $ \Gamma (0) =\Gamma_h$ and $\Gamma (|{\bf \pi}|) =\Gamma_e$.

Manifestations of the orbital order in FeSe have been seen in Raman, STM, ARPES, and other experiments
 (see Ref. \cite{recent} for recent review on FeSe). STM data analysis  within a single domain resolved one elliptical hole Fermi surface (FS)  and one electron FS, whose form becomes peanut-like below $T_s$ (Ref. \cite{davis}).
 In the 1FeBZ, where electron pockets are centered at $(0,\pi)$ and $(\pi,0)$, the observed hole pocket is elongated towards $(0,\pi)$, and the observed electron pocket is centered at $(\pi,0)$, and its  smaller axis is along the $Y$ direction (see Fig. \ref{fig:FS_Sketch}b).  ARPES data on single-domain samples~\cite{arpes_strained,watson,shibauchi,Zhou}
  show the same shape of the FSs.  In multi-domain samples, ARPES shows the combination of FSs from different domains~\cite{borisenko_latest}.
  In the tetragonal phase above $T_s$,  hole pockets are $C_4$ symmetric and electron pockets are elliptical (Fig. \ref{fig:FS_Sketch}a).   The changes from a $C_4$-symmetric to an elliptical shape for the hole pocket and  from an elliptical to a peanut-like shape for the $(\pi,0)$ electron pocket are due to orbital order.  Adding $\Gamma_h$ and $\Gamma_e$  terms to the hopping Hamiltonian  in orbital representation  and transforming from orbital to band basis, one obtains~\cite{davis,kang_latest} that the observed shapes of the pockets are reproduced if $\Gamma_h >0$ and $\Gamma_e <0$, i.e., the orbital order changes sign between hole and electron pockets.

\begin{figure}[t]
	\centering{}
	\subfigure[]{\includegraphics[width=0.48\columnwidth]{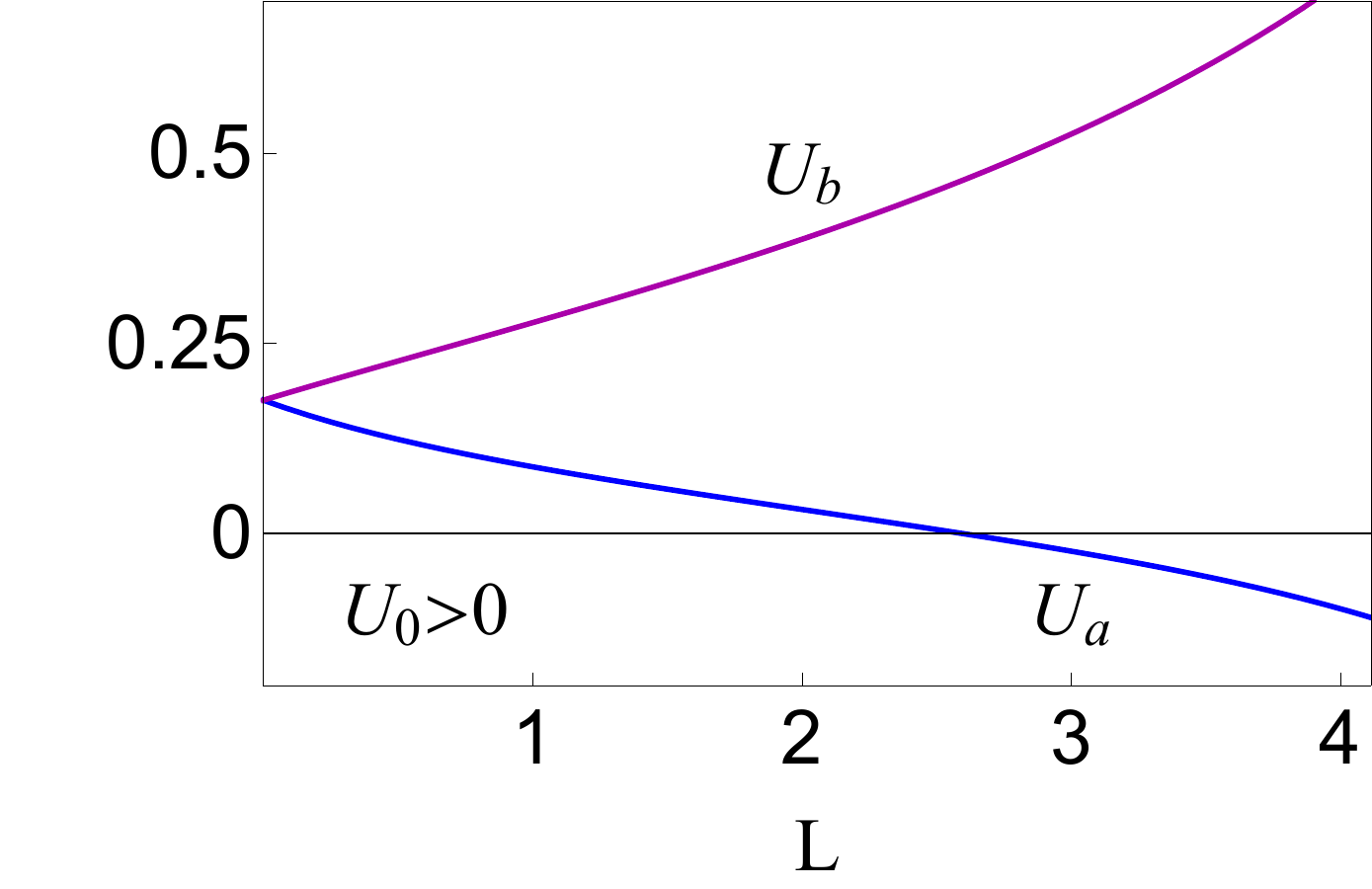}}
	\subfigure[]{\includegraphics[width=0.48\columnwidth]{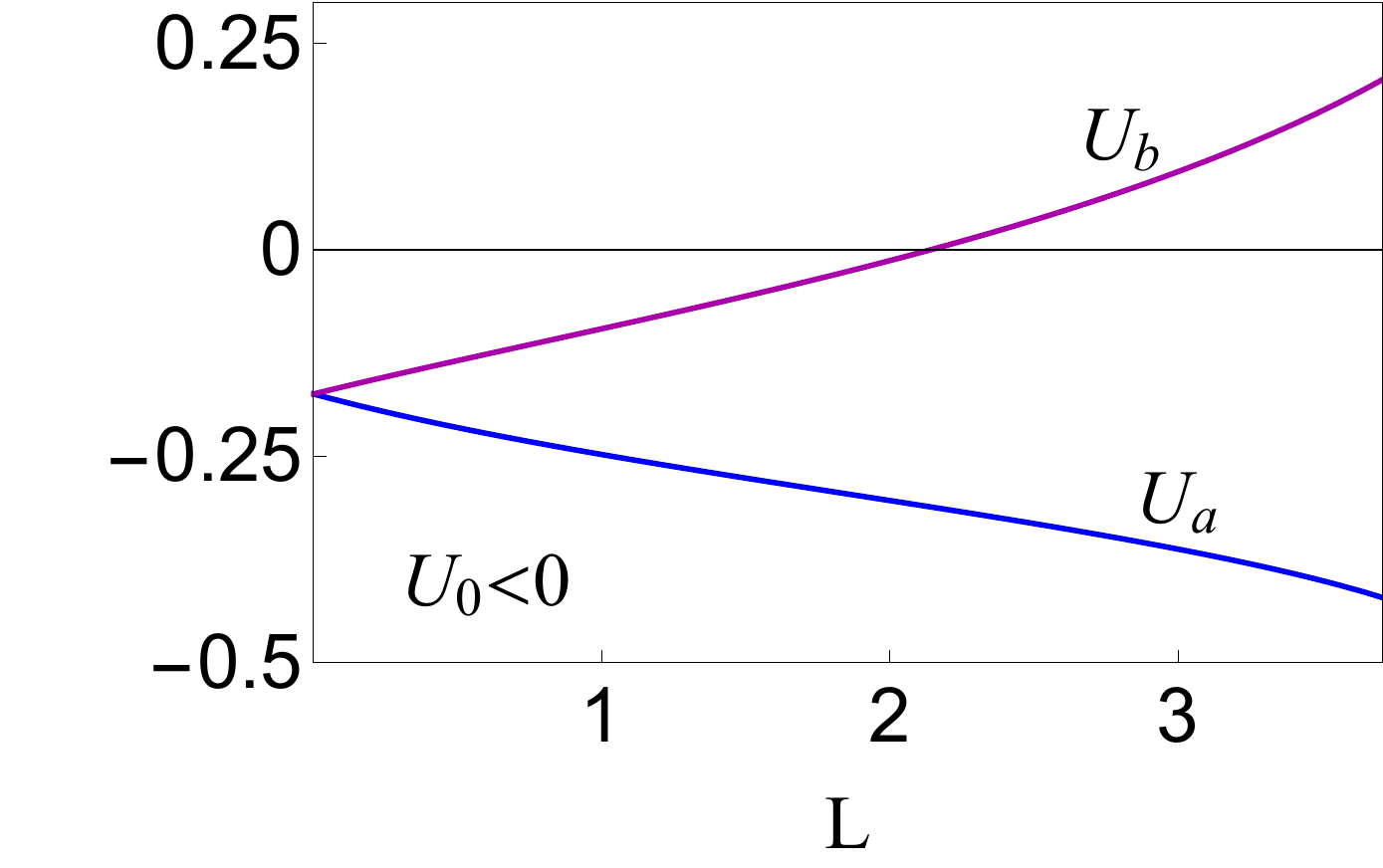}}
	\subfigure[]{\includegraphics[width=0.48\columnwidth]{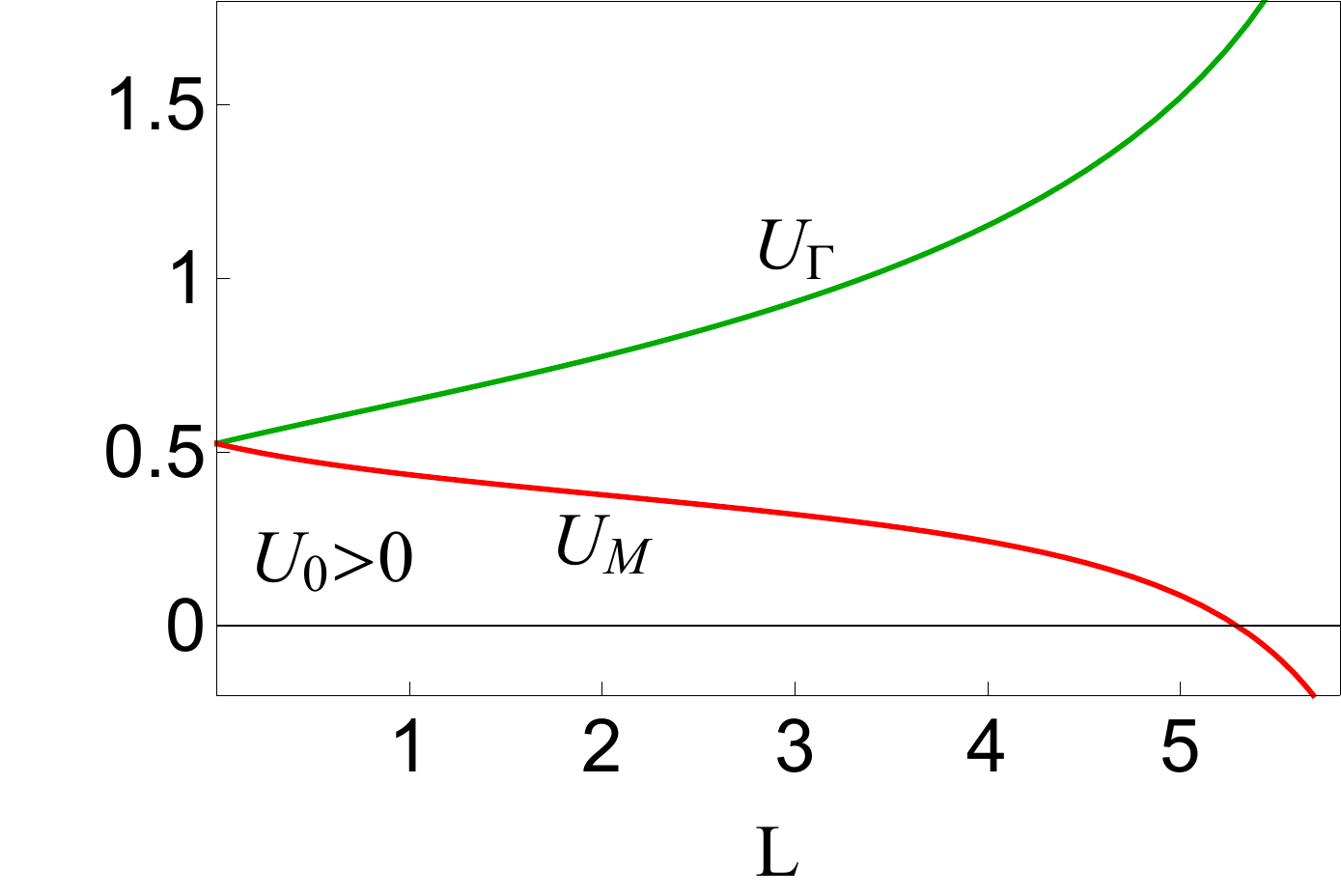}}
	\subfigure[]{\includegraphics[width=0.48\columnwidth]{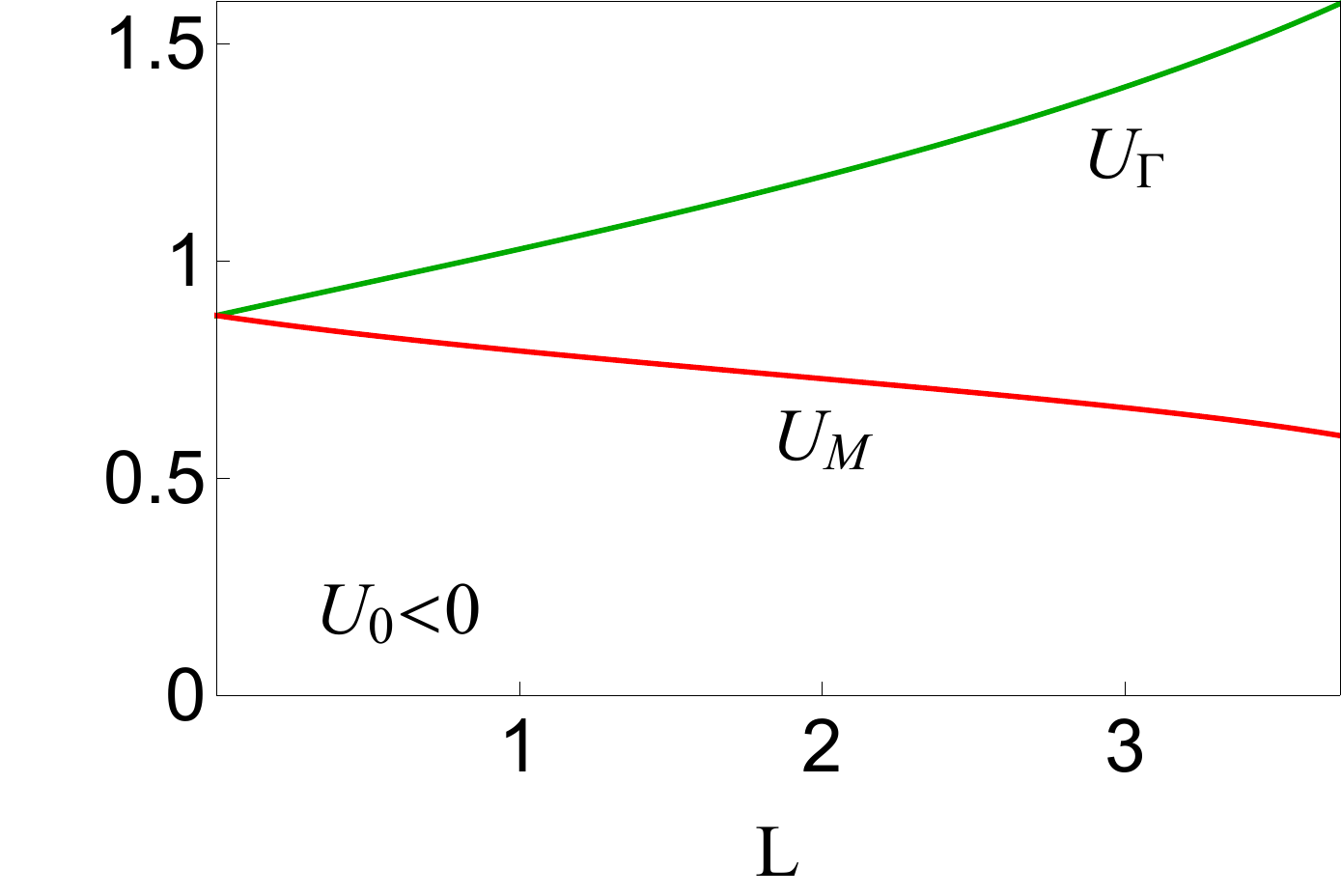}}
	\caption{Panels (a) and (b) -- the pRG flow of the couplings $U_a$ and $U_b$ for the case when the
 bare $U^{(0)}_a = U^{(0)}_b = U_0 = 5J-U$ is positive in (a) and negative in (b) (we set $J/U =0.3$ and $0.1$, respectively). Panels (c) and (d) -- the flow of the couplings
 $U_M$ and $U_{\Gamma}$ in Eq. (\ref{ch_3}).
 The parameter $L=log\frac{W}{E}$, where W is of order bandwidth and E is the running energy. The larger L is, the more high energy states are integrated out. We used
 $m_h U/(4\pi)=0.35$ where $m_h$ is the mass of the dispersion near the hole pocket.
 }\label{fig:interactions_flow}

	\end{figure}

\begin{figure}[t]
	\centering{}
	\subfigure[]{\includegraphics[width=0.48\columnwidth]{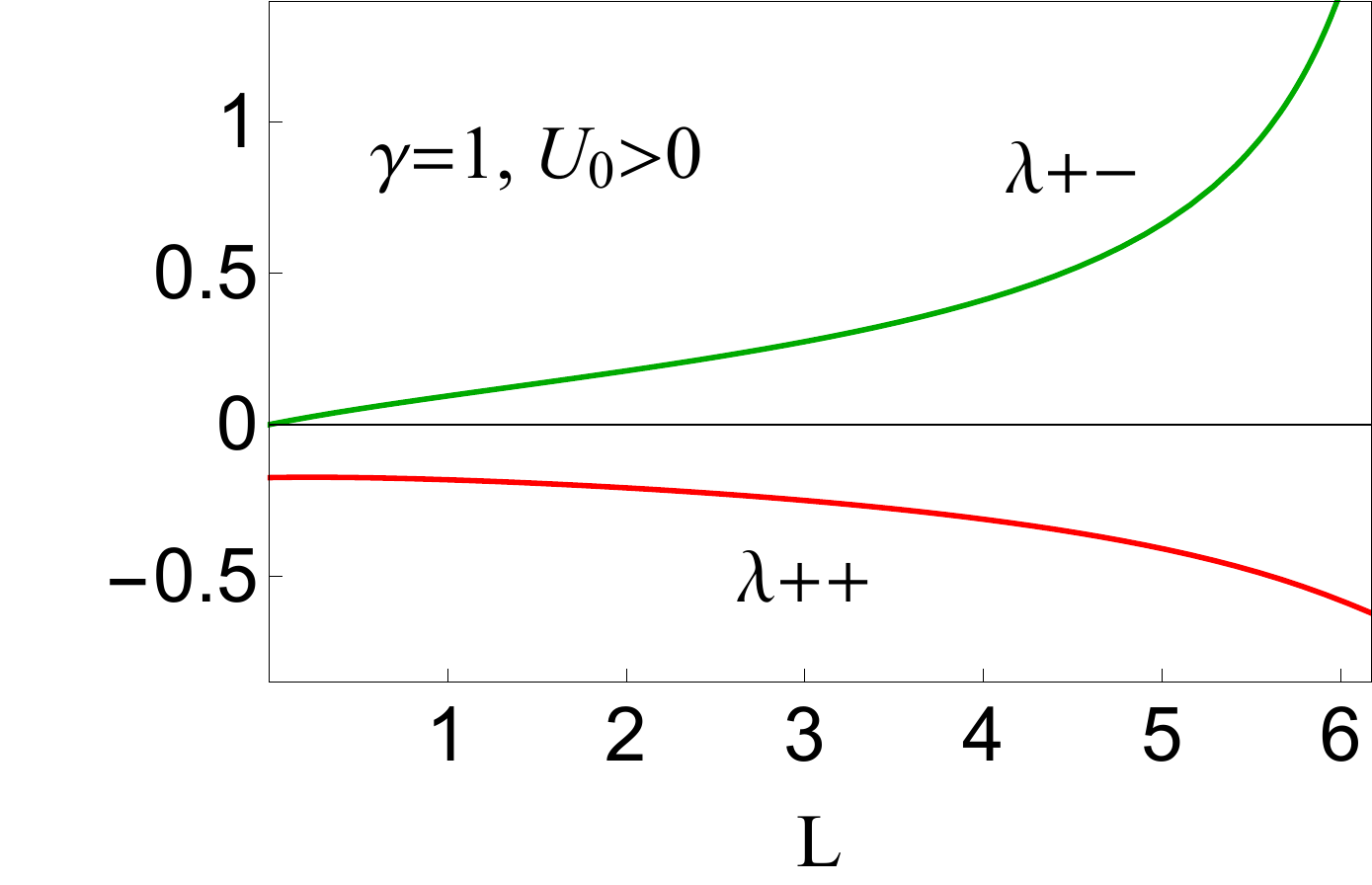}}
	\subfigure[]{\includegraphics[width=0.48\columnwidth]{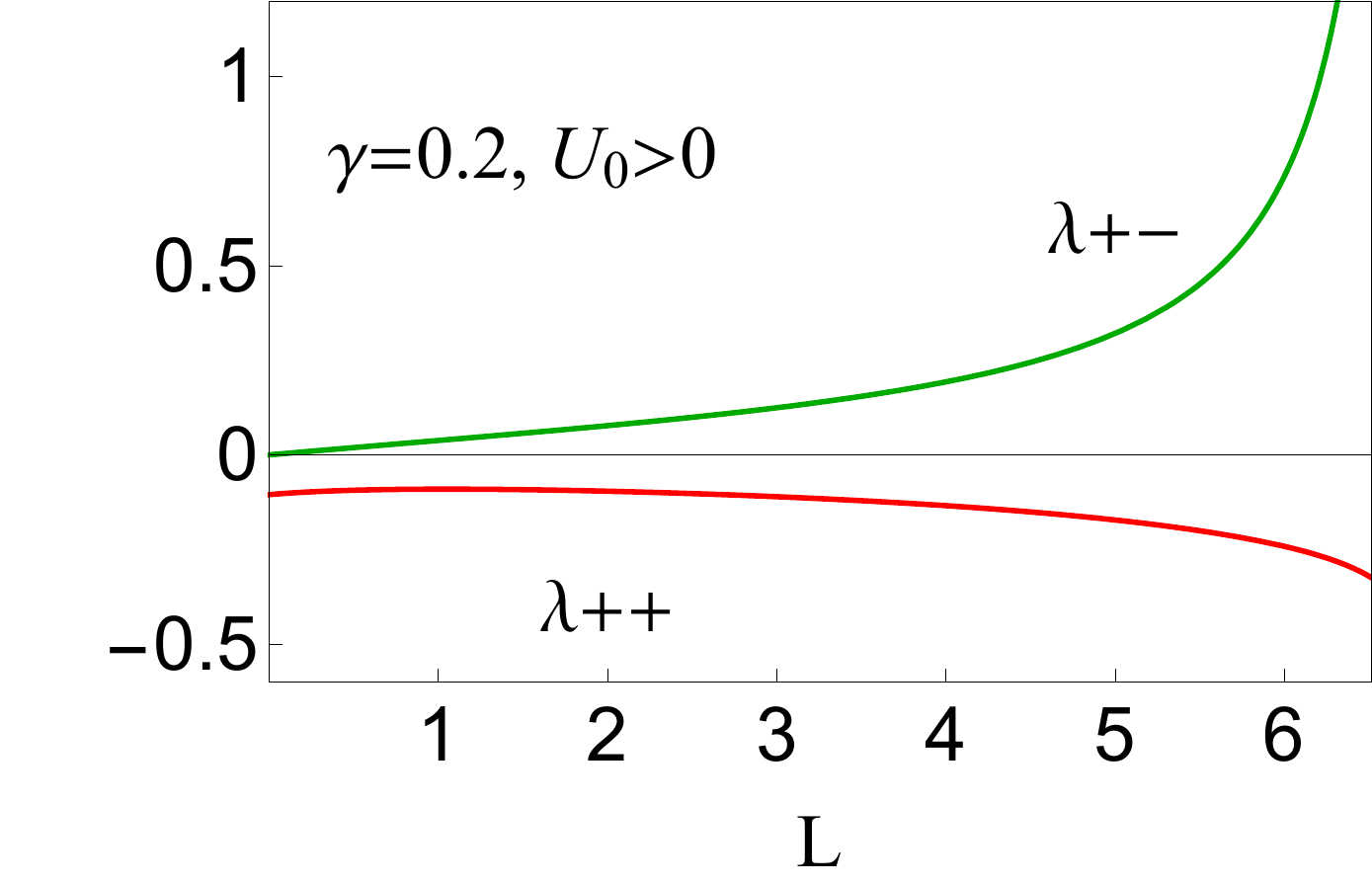}}
	\subfigure[]{\includegraphics[width=0.48\columnwidth]{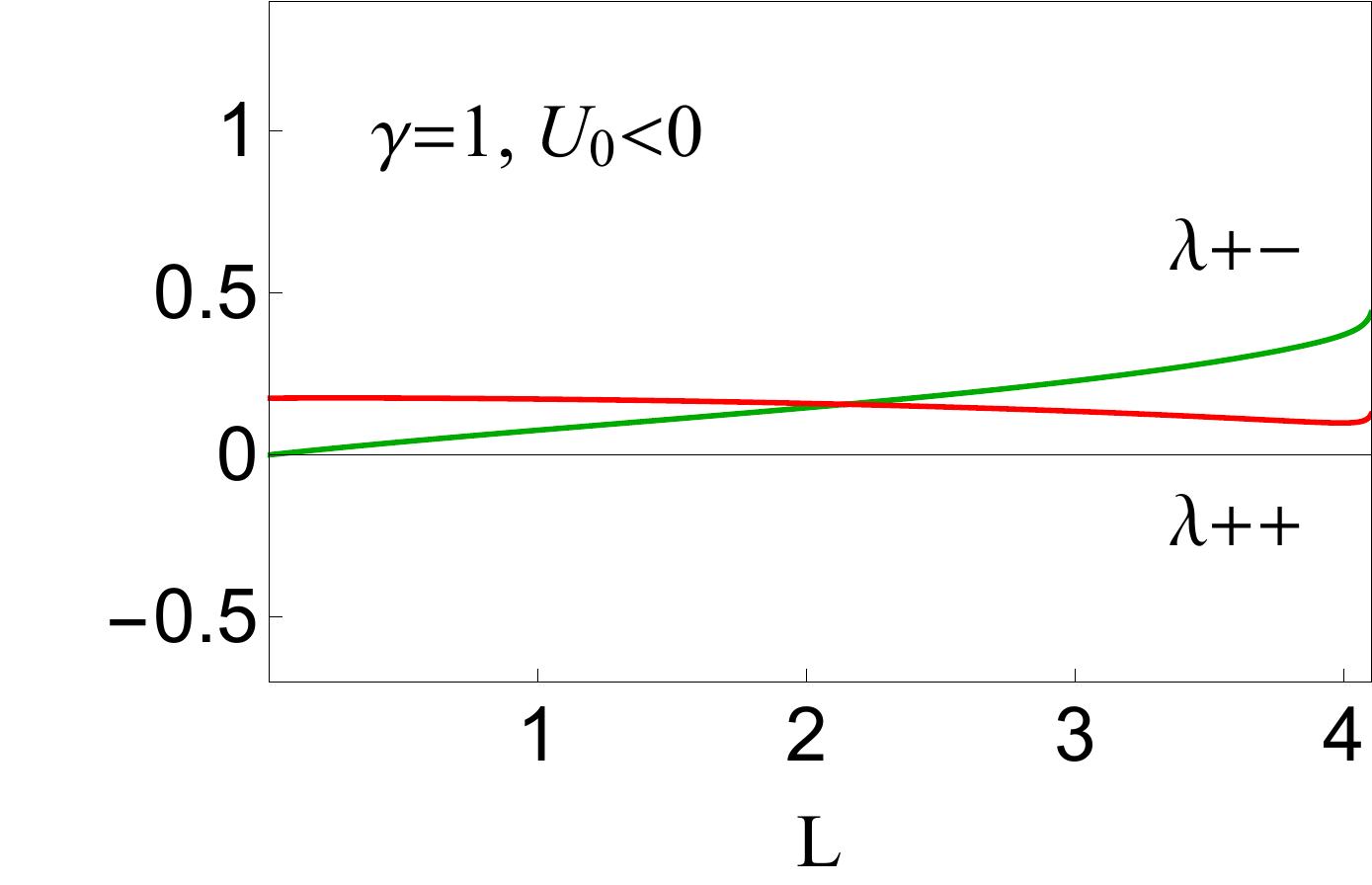}}
	\subfigure[]{\includegraphics[width=0.48\columnwidth]{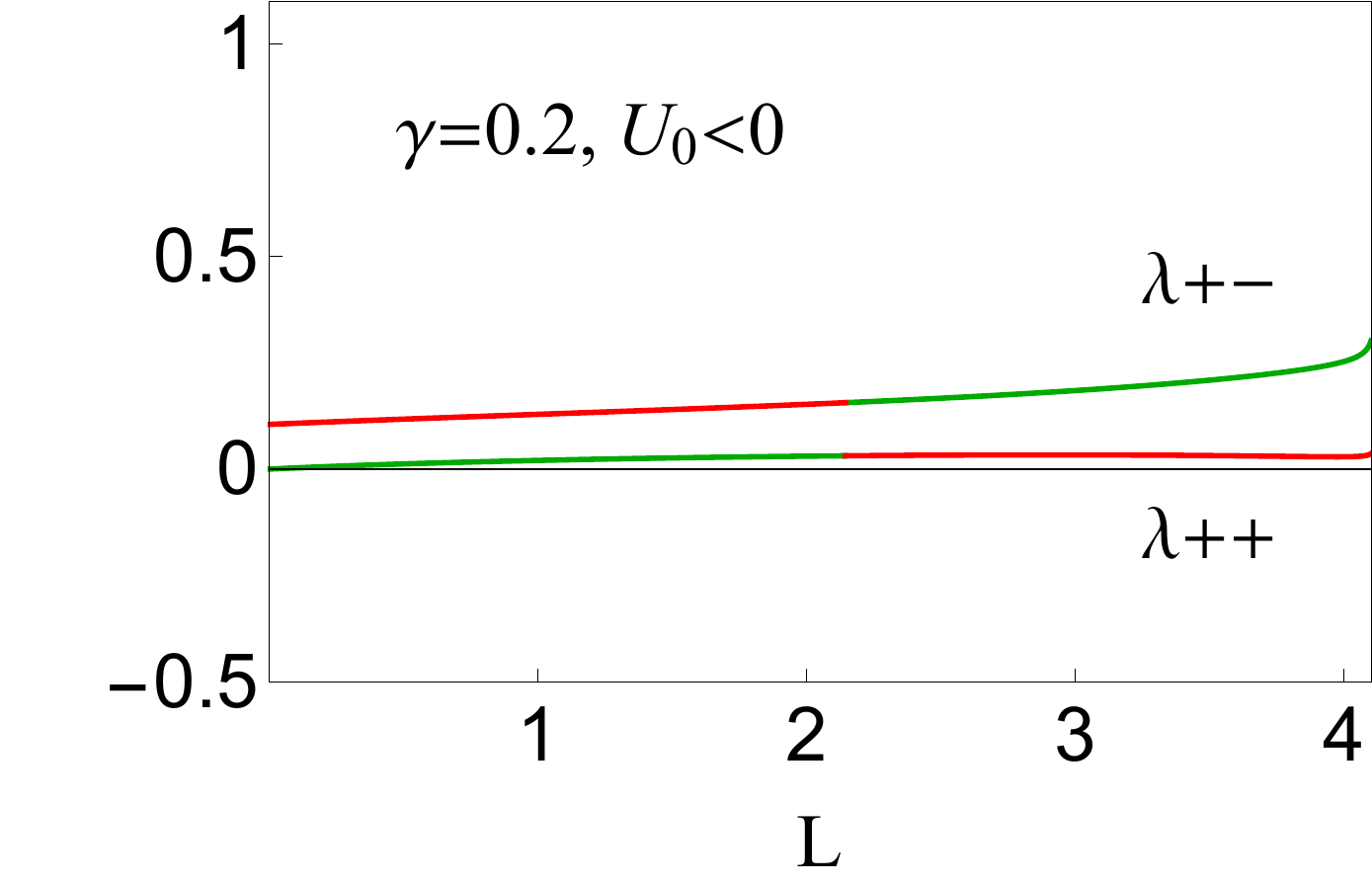}}
	\caption{The flow of the dimensionless couplings $\lambda^{++}$  in  sign-preserving $d^{++}$ channel (green)  and $\lambda^{+-}$ in sign-changing $d^{+-}$ channel (red). Notations are as in Fig. \protect\ref{fig:interactions_flow}.
Panels (a) and (b) - the flow for the case $U_0 = 5J -U >0$ for two values of the parameter $\gamma = A_e/A_h$ (see text). Panels (c) and (d)  -- the same for $U_0 <0$. The sign-changing $d^{+-}$ channel becomes dominant once $U_b$ changes sign near $L=2$. \label{fig:lambda}.
  For $\gamma \neq 1$, the couplings  jump by finite values when $U_b$ passes through zero. }
\end{figure}

  In this communication we address the issue how  the sign change between $\Gamma_e$ and $\Gamma_h$ can be understood theoretically.
  For this, we derive and analyze the self-consistent equation for  d-wave orbital order $\Gamma$.
  We argue that at mean-field level, the set of coupled equations for $\Gamma_h$ and $\Gamma_e$  contains the single effective interaction $U_{0}= 5J-U$,
    where $U$ and $J$ are Hubbard and Hund local interactions. The orbital order either does not develop, when $U_{0} >0$, or does develop, if $U_{0} <0$ and its magnitude is strong enough, but the solution necessarily yields equal sign of $\Gamma_e$ and $\Gamma_h$ ($d^{++}$ channel).
We next include into the analysis the fact that the couplings flow away from their bare values (used in mean-field analysis), when we progressively integrate out contributions of fermions with higher energies. This flow is captured within parquet renormalization group analysis (pRG)~\cite{Chubukov} or functional RG~\cite{RG}. The pRG flow splits $U_0$ into two different interactions $U_{a}$ and $U_{b}$.
   We show that this splitting gives rise to a non-zero coupling in another channel for orbital ordering, this time with $\Gamma_e$ and $\Gamma_h$ of opposite signs ($d^{+-}$ channel).   This is
   similar to how the coupling in the $s^{+-}$ pairing channel  emerges due to small inter-pocket pairing interaction
   on top of strong Hubbard repulsion. We show that the coupling  in this new orbital channel is attractive,
    regardless of
    the sign of the bare $U_{0}$, and exceeds the coupling in the $d^{++}$ channel.  Our results are summarized in Figs. \ref{fig:interactions_flow}, \ref{fig:lambda}.

  Our approach is similar to earlier works~\cite{ckf,kontani}, which also found an attraction in the $d^{+-}$ channel, but differs in detail.
  The authors of ~\cite{ckf} analyzed self-consistent equations for $\Gamma_h$ and $\Gamma_e$ in the $C_4$ symmetric regime using the values of the interactions $U_a$ and $U_b$ near the fixed trajectory, i.e., at the very end of the pRG flow.
  Here we consider the evolution of  $U_a$ and $U_b$ without assuming closeness to a fixed trajectory.  This is a more realistic approach, given that in practice pRG only runs over a finite window of energies.  We show that the $d^{+-}$ channel becomes attractive from the very beginning  of the pRG flow.
  The authors of ~\cite{kontani} considered the case of large $U/J$  and obtained sign-changing $d^{+-}$ orbital order by selecting a particular combination of RPA and Aslamazov-Larkin type diagrams for the renormalization of the Hubbard interaction.  We consider arbitrary $U/J$ and treat vertex renormalizations within pRG, which  accounts on equal footing for
   vertex renormalizations in particle-hole and particle-particle channels.
   Another explanation for the sign change between $\Gamma_e$ and $\Gamma_h$ has been put forward in Ref.~\cite{fanfarillo2016}. It is
   based on
    the earlier study\cite{ortenzi}, which showed that the self-energy due to spin fluctuation exchange has opposite sign near $\Gamma$ and near $X/Y$, and
     shrinks both hole and electron pockets.  The authors of ~\cite{fanfarillo2016} argued (on a semi-phenomenological level) that the
    X/Y anisotropy of spin fluctuations below $T_s$
    leads to $\text{sgn}(\Gamma_e)=-\text{sgn}(\Gamma_h)$.  Our approach is complimentary to that work: the authors of ~\cite{fanfarillo2016} included the X/Y anisotropy of the effective interaction but not orbital order.    We, on the contrary, include  orbital order into fermionic propagators, but neglect nematicity-induced changes of the interactions. We emphasize that both approaches lead to the sign change of the nematic splitting.

We also consider how orbital order affects the energies of $d_{xz}$ and $d_{yz}$ orbitals  $E_{xz}$ and $E_{yz}$ at $(0,\pi)$ and $(\pi,0)$ points in the 1FeBZ (the $M$ point in the 2FeBZ).  In absence of orbital order, the two energies are degenerate even in the presence of spin-orbit coupling\cite{Fernandes_Vafek}.  A non-zero $\Gamma_e$  breaks the degeneracy.  To first order in $\Gamma_e$, the energies
   split -- $E_{xz}$ increases by $\Gamma_e/2$ and $E_{yz}$ decreases by $\Gamma_e/2$.
  Observation of this splitting has been reported by Fedorov et al~\cite{Fedorov}. However, this group argued that the $d_{xz}/d_{yz}$ splitting appears on top of a larger effect -- a  simultaneous change of the temperature dependence of  $E_{xz}$ and $E_{yz}$ below $T_s$. According to Refs.~\cite{Fedorov,borisenko_latest}, both energies become smaller in magnitude.
    This observation is consistent with the later result by the same group~\cite{borisenko_latest}  that they can resolve both electron pockets  within a single domain, and both pockets have peanut-like form.
    A  simultaneous change of the temperature dependence of $E_{xz}$ and $E_{yz}$ below $T_s$  has also been reported in ~\cite{Coldea_AnnPhy}. Later, however,
     Watson et al. argued~\cite{watson} that they can only observe  $d_{yz}$ orbital at the $M$ point (in addition to $d_{xy}$). If this is the case, then the observed $T$ dependence below $T_s$ can be
    due to the expected first-order correction in $\Gamma_e$.
    To address this issue, we computed the corrections to  $E_{xz}$ and $E_{yz}$ to second order in $\Gamma_e$ and $\Gamma_h$.
    The $\Gamma^2_e$ and $\Gamma^2_h$  terms are the same for $E_{xz}$ and $E_{yz}$ and, if these terms are large, they can overtake the $\pm \Gamma_e/2$  splitting already at small $\Gamma_i$.
     We
     found that the
     second order
     contribution  accounts  only  for a  small correction to $\pm \Gamma_e/2$ and, moreover, the correction is of the wrong sign.
     If  both $E_{xz}$ and $E_{yz}$ indeed become smaller in magnitude below $T_s$, as the authors of Refs~\cite{Fedorov,borisenko_latest} argue,  this is
     due to some other physics than the one we consider here.
\begin{table}[h]
	\begin{center}
		\begin{tabular}{ | c | c | c || c | c | c || c | c | c |}
			\hline
			$\psi_i$ & Pocket & Orbital & $\psi_i$ & Pocket & Orbital & $\psi_i$ & Pocket & Orbital \\ \hline
			$\psi_1$ & Y & $d_{xz}$ & $\psi_3$ & X & $d_{yz}$ & $\psi_5$ & $\Gamma$ & $d_{yz}$\\ \hline
			$\psi_2$ & Y & $d_{xy}$ & $\psi_4$ & X & $d_{xy}$ & $\psi_6$ & $\Gamma$ & $d_{xz}$ \\ \hline
			\end{tabular}
	\end{center}
	\caption{\label{tab:affiliations}Affiliation of $\psi_i$ with a pocket and an orbital.}
\end{table}

\begin{figure}[t]
\subfigure[]{\includegraphics[width=0.99\columnwidth]{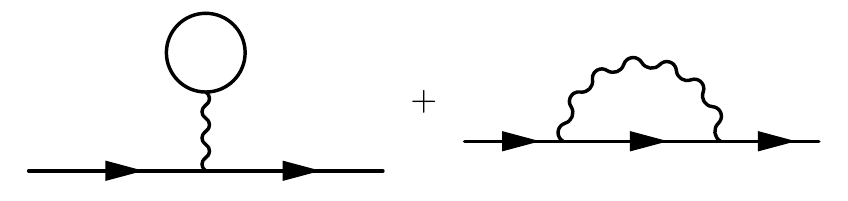}}\quad
\subfigure[]{\includegraphics[width=0.99\columnwidth]{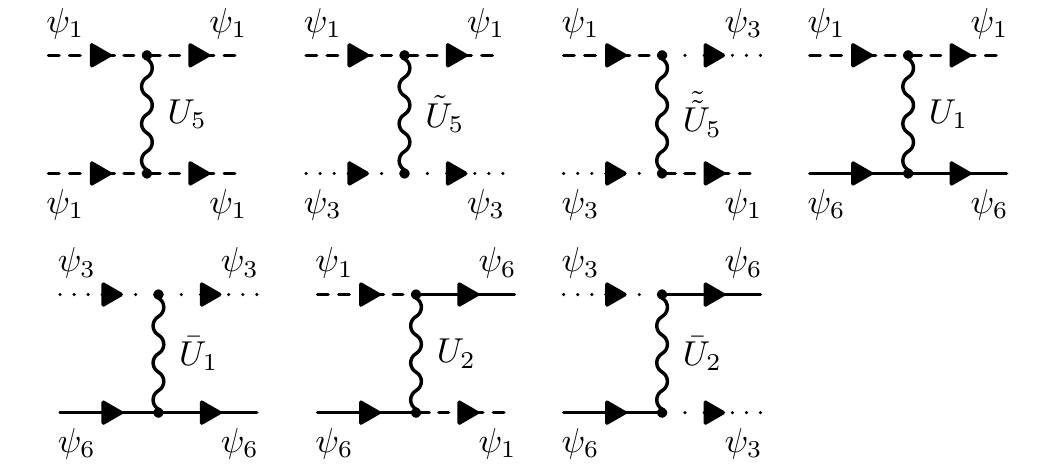}}
	\caption{a) Hartree and Fock self-energy diagrams;
     b) Examples of the interaction terms which contribute to Hartree-Fock self-energies.
     The $U_5$ terms in the first row also act on hole pockets ($\psi_5, \psi_6$). Each diagram has symmetry-equivalents.
     ($\psi_1\leftrightarrow \psi_3$, $\psi_5\leftrightarrow \psi_6$).
  The self-energy beyond mean-field has been computed using dressed interactions, which we obtained using pRG scheme. In a direct perturbation theory, this amounts to summing up infinite series of self-energy diagrams, including  RPA and Aslamazov-Larkin diagrams.  }
		\label{fig:Feynman_diagrams}
\end{figure}

{\it \bf Mean-field analysis}~~~
We consider a model  with two hole pocket near $(0,0)$ in the tetragonal phase (H-pockets) and
  two electron pockets near $(0,\pi)$ and $(\pi,0)$ in the 1FeBZ (Y and X pockets). The hole pockets are made out of $d_{xz}$ and $d_{yz}$ orbitals, the X pocket is made out of $d_{yz}$ and $d_{xy}$ orbitals and the Y pocket is made out of $d_{xz}$ and $d_{yz}$ (Refs. \cite{scalapino, cvetkovic}).
 We introduce six spices of fermions:  $\psi_1,\ldots,\psi_6$, see Tab~\ref{tab:affiliations} and two $d-$wave $d_{xz}/d_{yz}$ orbital order parameters
$\Gamma_h = \langle\psi^\dagger_6\psi_6-\psi^\dagger_5\psi_5\rangle$ and $\Gamma_e = \langle\psi^\dagger_1\psi_1-\psi^\dagger_3\psi_3\rangle $.
 For simplicity, we neglect d-wave orbital order on the $d_{xy}$ orbital (the $\psi^\dagger_2\psi_2-\psi^\dagger_4\psi_4$ term, Refs. \cite{Fernandes_Vafek,ruiqi}).
  At the mean-field level, the self-consistent equations for $\Gamma_h$ and $\Gamma_e$ are obtained by adding up Hartree and Fock diagrams for different orbitals (Fig. \ref{fig:Feynman_diagrams}a). To first order in the orbital order parameter, the self-energies are $\Sigma^H_{xz} = \Sigma_{h,0} + \Gamma_h/2$, $\Sigma^H_{yz} = \Sigma_{h,0} - \Gamma_h/2$,  $\Sigma^Y_{xz} = \Sigma_{e,0} + \Gamma_e/2$, $\Sigma^X_{yz} = \Sigma_{e,0} - \Gamma_e/2$, where
$\Sigma_{h,0}$ and $\Sigma_{e,0}$ are the self-energies in the absence of orbital order.  Evaluating the diagrams and taking the difference $\Sigma^H_{xz}-\Sigma^H_{yz} = \Gamma_h$,   $\Sigma^Y_{xz}-\Sigma^X_{yz} =\Gamma_e$, we obtain
 self-consistent equations for $\Gamma_h, \Gamma_e$ in the form~\cite{comm}
 \bea
&&\Gamma_h = U_a \left(n_{xz}^H -n_{yz}^H\right) + U_b \left(n_{xz}^Y -n_{yz}^X\right) \nonumber \\
&& \Gamma_e = U_a \left(n_{xz}^Y -n_{yz}^X\right) + U_b \left(n_{xz}^H -n_{yz}^H\right)
\label{ch_1}
\eea
Here each density $n_i$ is the momentum integral over the corresponding Fermi function.
 We find, to leading order in $\Gamma_i$,
$n_{xz}^H -n_{yz}^H = A_h \Gamma_h$ and
$n_{xz}^Y -n_{yz}^X  = A_e \Gamma_e$.  To obtain the prefactors $A_h$ and $A_e$, we used the orbitally-resolved low-energy model from Ref.~\cite{cvetkovic} for the kinetic energy, converted from orbital to band basis, and computed the momentum integrals of the Fermi functions for different bands, weighted by the "coherent" factors associated with the change of the basis.  We present the details of the calculations in~\cite{SM} and here state the result: both  $A_h$ and $A_e$ are negative, and their ratio $\gamma = A_e/A_h$ depends on the parameters in the kinetic energy and is, in general, of order one.
  Using the  band structure parameters that fit the ARPES and STM data, we obtained  $\gamma \sim 0.2$ (see~\cite{SM}).

The interactions $U_a$ and $U_b$ are  linear combinations of seven different interactions involving $d_{xz}$ and $d_{yz}$ orbital states near momenta where FSs are located.
 We show these seven interactions in Fig. \ref{fig:Feynman_diagrams}b.   In terms of these interactions,  $U_a = U_5 -2 {\tilde U}_5 + {\tilde {\tilde U}_5}$\cite{footnote}
 and $U_b = 2(U_1 - {\bar U}_1) - (U_2- {\bar U}_2)$ (labels are as in Fig. \ref{fig:Feynman_diagrams}b).   The bare values of the seven couplings are $U^{(0)}_5=U^{(0)}_1=U^{(0)}_2 = U, {\tilde U}^{(0)}_5 = {\bar U}^{(0)}_1 = U', {\tilde{\tilde U}^{(0)}_5} = {\bar U}^{(0)}_2 =J$.
 As a result, the bare $U^{(0)}_a$ and $U^{(0)}_b$ are the same: $U^{(0)}_a = U^{(0)}_b = U_0
 = U+J-2 U'$.  If we take $U'=U-2J$ (Ref. \cite{scalapino}), we obtain $U_0= 5J-U$.  Substituting $U_a=U_b = U_0$ into (\ref{ch_1}), we
 obtain that the only possible solution of the self-consistent set is $\Gamma_h=\Gamma_e$ (sign-preserving $d^{++}$ orbital order),  and this order develops if the eigenvalue $\lambda^{++} = U_0 (A_h + A_e) >1$.  The solution with the opposite sign of $\Gamma_e$ and $\Gamma_h$ does not emerge at the mean-field level.

{\it \bf Beyond mean-field}~~~
 We now go beyond mean-field and include into consideration
 that the seven interactions, which contribute to  $U_a$ and $U_b$,
  flow to different values as one progressively integrates out fermions with higher energies. This flow can be captured within pRG and
 comes from mutual vertex renormalizations of the total of 30 different interactions between low-energy fermions on $d_{xz}, d_{yz}$, and $d_{xy}$  orbitals~\cite{cvetkovic,ckf,ruiqi}.  The flow equations have been derived in ~\cite{ruiqi}, and we use the results of that work to obtain the flow of $U_a$ and $U_b$.  The results are shown in Fig.\ref{fig:interactions_flow}. We see that $U_a$ and $U_b$ become different from $U_0$, and $U_b > U_a$,
 irrespective of whether $U_0 >0$ or $U_0 <0$.  Solving  for the eigenfunctions and eigenvalues  of Eq. (\ref{ch_1}) when $U_a$ and $U_b$ are different, we obtain an eigenfunction $\Gamma^{++} = \Gamma_h + \alpha_{+} \Gamma_e$ with the
  eigenvalue $\lambda^{++}$ and   $\Gamma^{+-} = \Gamma_h + \alpha_{-} \Gamma_e$ with
  the eigenvalue $\lambda^{+-}$, where
  \bea
  &&\alpha_{\pm} = -\frac{1-\gamma}{2}\frac{U_a}{U_b} \pm \sqrt{\left( \frac{1-\gamma}{2} \right)^2\frac{U_a^2}{U_b^2} + \gamma } \label{ch_2} \\
  &&\lambda^{++,+-} =- |A_h| \left[ \frac{ 1+\gamma}{2}U_a \pm  U_b \sqrt{\left( \frac{1-\gamma}{2}\right)^2\frac{U_a^2}{U_b^2} + \gamma } \right] \nonumber
\eea

 We see that $\alpha_+ >0$ and $\alpha_- <0$, i.e., the eigenfunction $\Gamma^{++}$ describes  sign-preserving $d^{++}$ orbital order and $\Gamma^{+-}$ describes sign-changing $d^{+-}$ order.  We plot the corresponding eigenvalues $\lambda^{++}$ and $\lambda^{+-}$ in Fig. \ref{fig:lambda}.
  We see that $\lambda^{+-}$ becomes positive (i.e., attractive)  for any sign of $U_0$,  once we include the pRG flow of the interactions. We emphasize that this holds  even if the flow runs only over a small range of energies. For an instability towards a sign-changing orbital order, the flow needs to run over a finite range of energies to reach  $\lambda^{+-} >1$.

  For $U_0 >0$, the coupling in the $\lambda^{++}$ channel is repulsive, i.e., $d^{+-}$ orbital order is the only solution of Eq.~(\ref{ch_1}). For $U_0 <0$, the $d^{++}$ channel is
  attractive at the bare level, but we see from Fig. \ref{fig:lambda}c,d that it
  becomes
  sub-leading
  once $U_b$ changes sign under pRG (see Fig. \ref{fig:interactions_flow}b).
  The attraction in $d^{+-}$  orbital channel for $U_0<0$ was earlier obtained in Ref. \cite{kontani}   who used a combination of RPA spin and charge channels and Aslamazov-Larkin diagrams to separate $U_a$ and $U_b$.
In distinction with Ref.~\cite{kontani},  here  we account for the renormalization of $U_a$ and $U_b$ systematically, in an order-by-order treatment (as pRG is),
through all channels including the pairing channel.
  Like we said, we found that $\lambda^{+-}$ becomes positive already at the very beginning of the pRG flow, when the renormalization of $U_{a,b}$ can be obtained within a direct perturbative expansion. In particular, the condition $U_0<0$ is not required~\cite{comm_1}.
   We note in this regard that our computation of the self-energy, using the diagrams in Fig. \ref{fig:Feynman_diagrams}a with the dressed interactions obtained within the pRG scheme, is diagrammatically equivalent to summing up infinite series of contributions to the self-energy, including both RPA and Aslamazov-Larkin diagrams.

{\it \bf Temperature variations of $E_{xz}$ and $E_{yz}$.}~~~ We now analyze how the energies
$E_{xz}$ at $(0,\pi)$ and $E_{yz}$ at $(\pi,0)$ vary with increasing orbital order (in the 2FBZ these are energies of $d_{xz}/d_{yz}$ orbitals at $M$).   To first order in $\Gamma_e$, the two energies just split:
$E_{xz} = E_{e,0} + \Gamma_e/2$ and $E_{yz} =  E_{e,0} - \Gamma_e/2$, where $E_{e,0} <0$ is the energy in the absence of the nematic order~\cite{cvetkovic,Fernandes_Vafek}.  Our goal is to go beyond the first order in $\Gamma_e$ and check if  there is a large common term of order
 $\Gamma_{e,h}^2$.  A large positive $\Gamma_{e,h}^2$ term would be consistent with Refs. \cite{Fedorov,borisenko_latest}. The authors of these papers argued, based on interpretation of their ARPES data,
 that the magnitude of both $E_{xz}$ and $E_{yz}$ are reduced in the nematic phase.

To check this possibility, we computed the self-energies $\Sigma^Y_{xz}$ and $\Sigma^X_{yz}$ to order $\Gamma^2$. We
We did not do the  full self-consistent calculation to this order, as it would require to include the self-energy to order $\Gamma^2$ into the densities $n^H_{xz}, n^H_{yz}, n^Y_{xz}$, and $n^X_{yz}$. Rather, we evaluated the "source" term in the self-energy $\Sigma_{so} (\Gamma)$,  which comes from keeping $O(\Gamma_{h,e})$ terms in the self-energy, but expanding the densities to order $\Gamma^2_{h,e}$.  The
the common self-energy for $\Sigma^Y_{xz}$ and $\Sigma^X_{yz}$ below the nematic transition is proportional to $\Sigma_{so}(\Gamma) - \Sigma_{so} (0)$.
 We find
\beq
\Sigma_{so} (\Gamma) =  U_M (n^Y_{xz} + n^X_{yz}) + U_{\Gamma} (n^H_{xz} + n^H_{yz})
\label{ch_3}
\eeq
and $U_M = U_5 + 2 {\tilde U}_5 - {\tilde{\tilde U}_5}$ and $U_{\Gamma} = 2 (U_1 + {\bar U}_1) - (U_2 + {\bar U}_2)$.
 The bare value of
 $U_M$ and $U_{\Gamma}$ are again equal, each is $U + 2 U'-J$ ($=3U-5J$ if $U' = U-2J$), but  under pRG, $U_{\Gamma}$ becomes larger than $U_M$, as we show in Fig. \ref{fig:interactions_flow} c,d.
   The common densities are
 $(n^H_{xz} + n^H_{yz}) = n_{h,0} + B_h \Gamma^2_h,   (n^Y_{xz} + n^X_{yz}) = n_{e,0} + B_e \Gamma^2_e$,
 where $n_{i,0}$ labels the density for $\Gamma_i=0$.
 We find (see ~\cite{SM} for details)
 $B_h <0$ and $|B_e| \leq |B_h|$.
 In this situation, the common correction  to $E^Y_{xz}$ and $E^X_{yz}$ is negative. Given that $E_{e,0}$ is also negative, we see the common self-energy makes the two energies more negative.
  Furthermore,  the magnitude of   $B_h$ is at most of order $1/T_s$, hence
  near $T_s$, when $\Gamma_{h,e}$ are small, the self-energy to second order in $\Gamma_i$  is a small correction to the first-order $\pm \Gamma_i/2$ term.
  This is inconsistent with the interpretation of the data in Refs. \cite{Fedorov,borisenko_latest}.

{\it \bf Conclusions.} ~~~ In this communication we presented the solution of self-consistent equations for d-wave nematic order parameters
   on $d_{xz}/d_{yz}$ orbitals. We argued that at a mean-field level the only solution possible is sign-preserving $d^{++}$ nematic order $\Gamma$ (same sign of $\Gamma_e$ and $\Gamma_h$) when the bare coupling $U_0 <0$.  We went beyond mean-field and included the flow of the couplings under pRG. Then we found an attraction in $d^{+-}$ channel for  which $\Gamma_e$ and $\Gamma_h$ have opposite sign,
    in agreement with STM and ARPES data. We argued that $d^{+-}$ orbital order becomes the leading instability for either sign of bare $U_0$.   We also computed the common self-energy for $d_{xz}$ and $d_{yz}$ orbitals at the center of electron pockets to second order in $\Gamma$ to check whether we can reproduce the results of Refs. \cite{Fedorov,borisenko_latest} that the energies $E^Y_{xz}$ and $E^X_{yz}$ simultaneously get smaller by magnitude in the nematic phase.  We obtained a much smaller self-energy and of opposite sign  than the one which is needed.
     If the interpretation of the data in\cite{Fedorov,borisenko_latest} is correct, it has to be due to a
   self-energy with vertices beyond our RG analysis.

  {\it \bf Acknowledgments.} ~~~ We thank A. Coldea, L. Basconses, L. Benfatto, S. Borisenko, D. Chichinadze, R.M. Fernandes, J. Kang, T.K. Kim,  W. Ku, L. de' Medici, M. Watson, and Y.M. Wu for useful discussions. R.X. and A.V.C. are
 supported by the Office of Basic Energy Sciences, U.S. Department
 of Energy, under award DE-SC0014402.
 L.C. acknowledges support from the Alexander-von-Humboldt foundation.
 Work at BNL is supported by the U.S. Department of Energy (DOE), Division of Condensed Matter Physics and Materials Science, under Contract No. DE-SC0012704.  Part of the work was done while A.V.C. was visiting KITP in Santa Barbara. KITP is supported by NSF under Grant No. NSF PHY17-48958.   RX and LC equally contributed to this project. 

   \newpage

\begin{widetext}

\vspace{.3cm}
\begin{center}
\textbf{\large{}{}{}{}{}{}Supplemental Material \\ Orbital order in FeSe -- the case for vertex renormalization}{\large{}{}{}{}{}{}
}
\par\end{center}
\setcounter{equation}{0} \setcounter{figure}{0} \setcounter{table}{0}

\global\long\def\theequation{S\arabic{equation}}

\global\long\def\thetable{S\arabic{table}}
 \global\long\def\thefigure{S\arabic{figure}}




\section{Orbitally-resolved low-energy model}

We use an effective orbitally-resolved, low-energy model to describe excitations  near the Fermi pockets in FeSe. The model can be  constructed by expanding the hopping integrals near the centers of hole and electron pockets \cite{cvetkovic}.
We will  work in the "theoretical" 1FeBZ, where the hole pockets are centered at $\Gamma = (0,0)$, and the electron pockets are centered at $X= (\pi,0)$ and $Y = (0,\pi)$.  The two $\Gamma$-centered pockets are made out of $d_{xz}$ and $d_{yz}$ orbitals, the X pocket is made out of $d_{yz}$ and $d_{xy}$ orbitals and the Y pocket is made out of $d_{xz}$ and
 $d_{xy}$ (Refs. \cite{scalapino,cvetkovic}),  as shown in Fig.~\ref{fig:FS_Sketch}.
  The dispersion in the physical 2FeBZ can be obtained by folding the 1FeBZ along its diagonal. Under the folding, the points Y and X both become $M = (\pi,\pi)$.

We  introduce six spices of fermions:  $\psi_1,\ldots,\psi_6$.  Fermions $\psi_1$ and $\psi_2$ describe $d_{xz}$ and $d_{xy}$ excitations near the electron pocket at $Y$, $\psi_3$ and $\psi_4$ describe  $d_{yz}$ and $d_{xy}$ excitations near the $X$ pocket, and $\psi_5$ and $\psi_6$  describe $d_{xz}$ and $d_{yz}$ excitations near the hole pockets at $\Gamma$. The assignment is summarized in Tab~\ref{tab:affiliations} and sketched in Fig.~\ref{fig:FS_Sketch}.

The quadratic Hamiltonian for states close to the hole pocket is
\begin{equation}\label{H_k}
H^\Gamma=\sum_{\bm{k},\sigma}
\begin{pmatrix}
\psi_{5\sigma}^{\dagger}(\bm{k}),\psi_{6\sigma}^{\dagger}(\bm{k})            	
\end{pmatrix}
 h_{\Gamma}(\bm{k})
\begin{pmatrix}
\psi_{5\sigma}(\bm{k})\\
\psi_{6\sigma}(\bm{k})
\end{pmatrix}
\end{equation}
with
 \begin{equation}\label{ch_1_1}
	h_{\Gamma}(\bm{k})=\begin{pmatrix}\epsilon_{h}-\frac{k^{2}}{2m_{h}}+\frac{b}{2}(k_{x}^{2}-k_{y}^{2}) -\Gamma_h & ck_{x}k_{y}\\
		ck_{x}k_{y} & \epsilon_{h}-\frac{k^{2}}{2m_{h}}-\frac{b}{2}(k_{x}^{2}-k_{y}^{2}) +\Gamma_h
	\end{pmatrix},
\end{equation}
In the tetragonal phase, $\Gamma_h=0$. In the orthorhombic phase at $T < T_s$, $\Gamma_h$  has a finite value.

To make the formulas more compact, we assume $b=c$.  Then in the tetragonal phase the hole pockets are circular, with energies $\epsilon_{c,d}^{tet}=\epsilon_h-(k^2/(2m_h))(1\pm b m_h)$. The band operators $c_k$ and $d_k$ of the inner and outer hole pocket are  related to
$\psi_5 (k)$ and $\psi_6 (k)$ by a rotation
 \beq
 c_k=-\sin\theta_\Gamma (k)  \psi_5 (k)+\cos\theta_\Gamma (k) \psi_6(k), ~~~
 d_k=\cos\theta_\Gamma (k) \psi_5 (k)+\sin\theta_\Gamma (k) \psi_6(k)
 \label{tu_1}
 \eeq
 where  $\theta_\Gamma (k)$ corresponds to the polar angle along a hole pocket,  counted from the $k_x$-axis.

In the orthorhombic (nematic) phase, the dispersion of hole-like excitations is altered to
\begin{align}\label{eq:Eh}
\epsilon_{c,d}^{nem}=\epsilon_h-k^2/(2m_h) \mp\sqrt{b^2k^4+4\Gamma_h^2-4bk\Gamma_h\cos\theta}.
\end{align}
The transformation to band basis can still be viewed as a rotation, like in Eq. (\ref{tu_1}), but the rotation angle $\phi_\Gamma (k)$ no conger coincides with
$\theta_\Gamma (k)$ and is expressed as \cite{review_ch_fern,kang_latest}
\begin{align}
\tan{\phi_\Gamma (k)} = \frac{\sin{2\theta_\Gamma (k)}}{\cos{2\theta_\Gamma (k)} - 2\Gamma_h/(b k^2)}
\end{align}

 The excitations near the electron pockets are described by
\begin{equation}
H^{Y,X}=\sum_{\bm{k},\sigma}
\begin{pmatrix}
\psi_{1,3\sigma}^{\dagger}(\bm{k}),\psi_{2,4\sigma}^{\dagger}(\bm{k})            	
\end{pmatrix}
 h_{Y,X}(\bm{k})
\begin{pmatrix}
\psi_{1,3\sigma}(\bm{k})\\
\psi_{2,4\sigma}(\bm{k}).
\end{pmatrix}
\end{equation}
 with
 \begin{equation}
	h_{Y,X}(\bm{k})=\begin{pmatrix}\epsilon_{1}+\frac{k^{2}}{2m_{1}} \pm \frac{a_1}{2}(k_{x}^{2} - k_{y}^{2}) \pm \Gamma_e & -\sqrt{2}ivk_{x/y} \\
		\sqrt{2}ivk_{x/y} & \epsilon_{2} + \frac{k^{2}}{2m_{2}} \pm \frac{a_2}{2}(k_{x}^{2} - k_{y}^{2})
	\end{pmatrix}
\label{ch_1_2}
\end{equation}
where the upper sign is for the $Y$ pocket and the lower for the $X$ pocket.  In the tetragonal phase $\Gamma_e=0$, in the orthorhombic phase it is finite.
 The parameters $v, \epsilon_{1,2}$, $a_{1,2}$, and $m_{1,2}$ can be determined by fitting the band structure to ARPES data. For FeSe they are given in the supplemental of Ref.~\cite{kang_latest}.
 In the band basis, this gives two branches around the $X$ point and two branches around the $Y$ point.
    Only one dispersion from each pair  crosses the Fermi level and forms the electron pocket at $X$ or $Y$.

 In the tetragonal phase the energies of the bands that cross the Fermi level are given by $\epsilon_{X,Y}^{tet}=\frac{1}{2}(C_1^{X,Y}+C_2^{X,Y}) + \frac{1}{2}\sqrt{(C_1^{X,Y}-C_2^{X,Y})^2+ 8v^2k^2_{y,x}}$ with $C_i^{X,Y}=\epsilon_i + k^2/(2m_i)\mp a_i/2(k_x^2-k_y^2)$.
The bands that do not cross the Fermi level have energies $\tilde\epsilon_{X,Y}^{tet}=\frac{1}{2}(C_1^{X,Y}+C_2^{X,Y}) - \frac{1}{2}\sqrt{(C_1^{X,Y}-C_2^{X,Y})^2+ 8v^2k^2_{y,x}}$.
The conversion from orbital to band basis can be again written as rotation
\beq
e_{X,Y}=\mp i\cos\phi_{X,Y}\psi_{3,1}+\sin\phi_{X,Y}\psi_{4,2}, ~~~
\tilde e_{X,Y}=\pm i\sin\phi_{X,Y}\psi_{3,1}+\cos\phi_{X,Y}\psi_{4,2}
\label{tu_2}
\eeq
where $e_i$ labels the band that forms the electron pocket and $\tilde e_i$ the one that does not cross the Fermi level. However, the
 angle $\phi_{X,Y}$ does not coincide with the polar angle along an electron pocket.
To a good approximation, $\cos\phi_{X,Y}=A\sin\theta_{X,Y}$ and $\sin\phi_{X,Y}=\sqrt{1-A^2\sin^2\theta_{X,Y}}$, where $\theta_{X(Y)}$ is the polar angle measured along $\Gamma-X$ ($\Gamma-Y$) and $1/\sqrt{2}<A<1$ is a constant\cite{kang_latest,ruiqi}.

In the  orthorhombic phase, the energies of the bands that cross the Fermi level become
\begin{align}\label{eq:Ee}
\epsilon_{X,Y}^{nem}=\frac{1}{2}(C_1^{X,Y}+C_2^{X,Y}\mp\Gamma_e) + \frac{1}{2}\sqrt{(C_1^{X,Y}-C_2^{X,Y}\mp \Gamma_e)^2+ 8v^2k^2_{y,x}}.
\end{align}
The transformation from orbital to band space can still be written as in (\ref{tu_2}), but the relations between $\phi_{X,Y}$  and $\theta_{X,Y}$
change to $\cos\phi_{X,Y}=A_{X,Y}\sin\theta_{X,Y}$ and $\sin\phi_{X,Y}=\sqrt{1-A^2_{X,Y}\sin^2\theta_{X,Y}}$ with $A_{X,Y}=A(1\mp \frac{\Gamma_e}{\Delta E}(1-A^2\sin^2\theta_{X,Y}))$ and $\Delta E =\tilde \epsilon_X^{tet}(k_F^X)$,
where $k_F^X$ is the Fermi wave vector of the electron pocket at $X$ (Ref. \cite{kang_latest}). In the following, we approximate  $A_{X,Y}$ by their average values along the electron FSs:
$A_{X,Y} =A(1 \mp \beta \Gamma_e)$ with $\beta=(1-A^2/2)/\Delta E$. The constant $\Delta E$ is given by $\tilde \epsilon_X^0(k_F^X)$, where $k_F^X$ is the Fermi wave vector of the electron pocket at $X$.

 There are 30 symmetry-allowed interactions between  six fermion species $\psi_i$~\cite{cvetkovic,ruiqi}.
  We assume that interactions involving $d_{xy}$ fermions are small, by one reason or the other
  (e.g., by applying full pRG to the full  model with 30 couplings~\cite{ruiqi}),
  and focus on the interaction terms which involve  fermions from $d_{xz}$ and $d_{yz}$ orbitals. These interactions  are
\allowdisplaybreaks
\begin{align}\label{interaction}
H_{int}= & {U}_{1}\sum\nolimits'\left[\psi_{1\sigma}^{\dag}\psi_{1\sigma}\psi_{6\sigma'}^{\dag}\psi_{6\sigma'}+
\psi_{3\sigma}^{\dag}\psi_{3\sigma}\psi_{5\sigma'}^{\dag}\psi_{5\sigma'}\right]+
\bar{U}_{1}\sum\nolimits'\left[\psi_{1\sigma}^{\dag}\psi_{1\sigma}\psi_{5\sigma'}^{\dag}\psi_{5\sigma'}+
\psi_{3\sigma}^{\dag}\psi_{3\sigma}\psi_{6\sigma'}^{\dag}\psi_{6\sigma'}\right]\notag\\
+ & {U}_{2}\sum\nolimits'\left[\psi_{1\sigma}^{\dag}\psi_{6\sigma}\psi_{6\sigma'}^{\dag}\psi_{1\sigma'}+
\psi_{3\sigma}^{\dag}\psi_{5\sigma}\psi_{5\sigma'}^{\dag}\psi_{3\sigma'}\right]+
\bar{U}_{2}\sum\nolimits'\left[\psi_{1\sigma}^{\dag}\psi_{5\sigma}\psi_{5\sigma'}^{\dag}\psi_{1\sigma'}+
\psi_{3\sigma}^{\dag}\psi_{6\sigma}\psi_{6\sigma'}^{\dag}\psi_{3\sigma'}\right]\notag\\
+& \frac{U_{4}}{2}\sum\nolimits'\left[\psi_{5\sigma}^{\dag}\psi_{5\sigma}\psi_{5\sigma'}^{\dag}\psi_{5\sigma'}+\psi_{6\sigma}^{\dag}\psi_{6\sigma}\psi_{6\sigma'}^{\dag}\psi_{6\sigma'}\right]\notag\\
+ & \tilde{U}_{4}\sum\nolimits'\psi_{5\sigma}^{\dag}\psi_{5\sigma}\psi_{6\sigma'}^{\dag}\psi_{6\sigma'}+\tilde{\tilde{U}}_{4}\sum
\nolimits'\psi_{5\sigma}^{\dag}\psi_{6\sigma}\psi_{6\sigma'}^{\dag}\psi_{5\sigma'}+ \frac{U_{5}}{2}\sum\nolimits'\left[\psi_{1\sigma}^{\dag}\psi_{1\sigma}\psi_{1\sigma'}^{\dag}\psi_{1\sigma'}+\psi_{3\sigma}^{\dag}\psi_{3\sigma}\psi_{3\sigma'}^{\dag}\psi_{3\sigma'}\right]\notag\\
+&  \tilde{U}_{5}\sum\nolimits'\psi_{1\sigma}^{\dag}\psi_{1\sigma}\psi_{3\sigma'}^{\dag}\psi_{3\sigma'}+\tilde{\tilde{U}}_{5}\sum\nolimits'\psi_{1\sigma}^{\dag}\psi_{3\sigma}\psi_{3\sigma'}^{\dag}\psi_{1\sigma'}\notag\\
\end{align}
The summation is over spin components  and over momenta, under the constraint of momentum conservation.
These interactions describe all symmetry-allowed scattering processes between $d_{xz}$ and $d_{yz}$ fermions near electron and hole pockets.
We omitted pair hopping terms because they do not play a role in the following.

If we use the microscopic model with local Hubbard-Hund interactions,  the interaction parameters are~\cite{ruiqi}
 \begin{align}
{U}_{1} & ={U}_{2} =U_{4}=U_{5}
=U,\notag\\
\bar{U}_{1}&=\tilde{U}_{4}=\tilde{U}_{5}=U',\notag\\
\bar{U}_{2}&=\tilde{\tilde{U}}_{4}=\tilde{\tilde{U}}_{5}=J,\notag\\
\label{Hubbard_relation}
\end{align}

\section{Self-consistent equations for $\Gamma_h$ and $\Gamma_e$}
To obtain the self-consistent equations for the nematic order parameters $\Gamma_h$ and $\Gamma_e$, we evaluate the Hartree and Fock diagrams for the self-energies $\Sigma_{xz}$ and $\Sigma_{yz}$ near hole and electron pockets.  In orbital basis, these self-energies contribute to the diagonal terms in
 Eqs. (\ref{ch_1_1}) and (\ref{ch_1_2}).
  Each self-energy contains a piece from the tetragonal phase and a piece which depends on $\Gamma_e$ and $\Gamma_h$.
  To first order in the orbital order, $\Sigma_{xz}^\Gamma = \Sigma_{tet}^{\Gamma} + \Gamma_h/2$ and $\Sigma_{yz}^\Gamma = \Sigma_{tet}^{\Gamma} - \Gamma_h/2$, where $\Sigma_{tet}^\Gamma$ is the self-energy at the $\Gamma$ point in the tetragonal phase.  At momenta close, but not equal to $\Gamma$, the self-energy acquires some $k$-dependence already in the tetragonal phase, but this dependence is small and irrelevant for our purposes, and we neglect it.
  Taking the difference between the two self-energies, we obtain $\Sigma_{xz}^\Gamma -\Sigma_{yz}^\Gamma=\Gamma_h$.
   Similarly, $\Sigma_{xz}^Y-\Sigma_{yz}^X=\Gamma_e$.
     Evaluating the diagrams for the self-energies then leads to
 \bea
&&\Gamma_h = U_{a4} \left(n_{xz}^\Gamma -n_{yz}^\Gamma\right) + U_b \left(n_{xz}^Y -n_{yz}^X\right) \nonumber \\
&& \Gamma_e = U_{a5} \left(n_{xz}^Y -n_{yz}^X\right) + U_b \left(n_{xz}^\Gamma -n_{yz}^\Gamma\right),
\label{ch_3_1}
\eea
The differences $n_{xz}^\Gamma -n_{yz}^\Gamma$ and $n_{xz}^Y -n_{yz}^X$ vanish in the tetragonal phase and are proportional to $\Gamma_h$ and $\Gamma_e$, respectively. Eq. (\ref{ch_3_1}) then becomes a self-consistent set of linarized equations for $\Gamma_h$ and $\Gamma_e$. Solving the set, one obtains $T_s$ and the ratio $\Gamma_e/\Gamma_h$ near $T_s$. We assume that the sign of $\Gamma_e/\Gamma_h$ will not change at a smaller $T$, when non-linear terms in the r.h.s. of (\ref{ch_3_1}) become relevant.

The interaction terms in (\ref{ch_3_1}) are  $U_{a5} = U_5 -2 {\tilde U}_5 + {\tilde {\tilde U}_5}$, $U_{a4} = U_4 -2 {\tilde U}_4 + {\tilde {\tilde U}_4}$ and $U_b = 2(U_1 - {\bar U}_1) - (U_2- {\bar U}_2)$.
 If we use the local Hubbard-Hund model Eq.~\eqref{Hubbard_relation},
 we obtain $U_{a5} = U_{a4} = U_b = U+J - 2U'$. However, the couplings become different once we include vertex corrections to $U_{a5}, U_{a4}$, and $U_b$.
  In  the main text
 we present the results for the dressed couplings assuming that the running  $U_{a5}$ and  $U_{a4}$ remain equal, i.e., the running  $U_{a5} = U_{a4} = U_a$.
 Here we present the results for a generic case when only bare $U_{a5} = U_{a4}$, but the running couplings are different. The running couplings $U_{a5}$ and $U_{a4}$ follow each other, and their ratio  tends to constant $r$, whose value depends on system parameters~\cite{ruiqi}.  Below we use $U_{a5} = U_a$ and $U_{a4} = r U_a$, when we will be using the dressed couplings.

The fermionic densities are the integrals over momentum and sums over Matsubara frequencies  of the corresponding Green's functions: $(n_{xz}^\Gamma -n_{yz}^\Gamma)=T\sum_{\omega_m}\int d^k/(2\pi)^2(G_{xz}^ \Gamma (k,\omega_m)-G_{yz}^\Gamma (k, \omega_m))$ and $(n_{xz}^Y -n_{yz}^X)=T\sum_{\omega_m}\int d^k/(2\pi)^2(G_{xz}^Y (k,\omega_m)-G_{yz}^X (k,\omega_m))$.
To evaluate the integrals, we transfer the Green's functions to the band basis and express the result in terms of the corresponding Fermi functions. One can check that in the tetragonal phase the self-energies
  $\Gamma_{tet}^\Gamma$ and $\Gamma_{tet}^{X,Y}$  come from states not confined to the FSs, however the additional terms in the orthorhombic phase, proportional to $\Gamma_h$ and $\Gamma_e$, come from the states near $\Gamma$, X, or Y points.
In explicit form, we obtain near hole pockets
\begin{align}\label{eq:nh}
&(n_{xz}^\Gamma -n_{yz}^\Gamma)=\int \frac{d^2k}{(2\pi)^2} \cos 2\phi_H \Big( n_F(\epsilon_c^{nem}) - n_F(\epsilon_d^{nem}) \Big) = A_h \Gamma_h \nonumber \\
&A_h =- \int \frac{d^2k}{(2\pi)^2} \left[\frac{2}{bk_F^2}\sin^2 2\theta \Big( n_F(\epsilon_c^{tet}) - n_F(\epsilon_d^{tet}) \Big) +\frac{1}{T}\cos^2 2\theta \left(\frac{e^{\epsilon_c^{tet}/T}}{\left(1+e^{\epsilon_c^{tet}/T}\right)^2} + \frac{e^{\epsilon_d^{tet}/T}}{\left(1+e^{\epsilon_d^{tet}/T}\right)^2}\right) \right] + \mathcal{O}(\Gamma_h^2)
\end{align}
 where $n_F$ is the Fermi distribution function and the expressions for $\epsilon_c^{tet} = \epsilon_c^{tet} (k) $ and $\epsilon_d^{tet} = \epsilon_d^{tet} (k) $ are given above.
 We recall that $c-$operators describe the inner hole pocket and $d-$operators describe the outer hole pocket. Accordingly,
  $\epsilon_c^{tet}\leq\epsilon_d^{tet}$,  so that $n_F(\epsilon_c^{tet}) \geq n_F(\epsilon_d^{tet}$).  As a result, both terms in the last line in (\ref{eq:nh}) have the same sign, hence $A_h <0$. We used this in the main text.

Similarly, fermionic densities in the vicinity of the electron pockets are given by
\begin{align}\label{eq:ne}
(n_{xz}^Y -n_{yz}^X)&=\int \frac{d^2k}{(2\pi)^2} \left[\cos^2\phi_Y n_F(\epsilon_Y) -\cos^2\phi_X n_F(\epsilon_X)\right] \nonumber \\
&= \int \frac{d^2k}{(2\pi)^2} \left[ A^2(1+\beta \Gamma_e)^2\sin^2\theta_Y n_F(\epsilon^{nem}_Y) - A^2(1-\beta \Gamma_e)^2\sin^2\theta_X n_F(\epsilon^{nem}_X)\right] = A_e \Gamma_e\nonumber \\
&A_e = A^2  \int \frac{d^2k}{(2\pi)^2} \cos^2\theta \left[ 2\beta n_F(\epsilon^{tet}_Y) - \frac{1}{T}f_Y(k) \frac{e^{\epsilon^{tet}_Y/T}}{\left(1+ e^{\epsilon^{tet}_Y/T} \right)^2} \right] + \mathcal{O}(\Gamma_e^2).
\end{align}
Here, the function $f_Y(k)$ is obtained when expanding the band energies to linear order in $\Gamma_e$, e.g., $\epsilon_Y^{nem}=\epsilon_Y^{tet}+ f_Y(k)\Gamma_e/2$ with $f_Y(k)=1+(C_1^Y-C_2^Y)/\sqrt{(C_1^Y-C_2^Y)^2+8v^2k_x^2}$.
To check the sign of $A_e$ we evaluate the momentum integral in the last line in (\ref{eq:ne}) analytically by setting $a_{1,2} = v=0$ in (\ref{ch_1_2}).  We obtain
\be
A_e=A^2 \frac{m_e}{2\pi}\left( \beta |\epsilon_1| + \beta T \ln \left( 1+e^{-|\epsilon_1|/T} \right) -\frac{1}{1+e^{-|\epsilon_1|/T}} \right)
\ee
For FeSe, $\beta|\epsilon_1|\sim1/4$. For such $\beta$, $A_e$ is negative.
 We then determine $A_e$ numerically, using the full quadratic Hamltonian in (\ref{ch_1_2}) and the values of  parameters for FeSe as given in Ref.~\cite{kang_latest}.  We again obtain that $A_e$ is negative.
 We used that $A_e<0$ in the main text. The magnitudes of $A_h$ and $A_e$ are comparable, but $A_h$ is larger: we found numerically
 $\gamma=A_e/A_h\approx 0.2$.

Using these results, we can write the self-consistent equations on $\Gamma_h$ and $\Gamma_e$ as
 \bea \label{eq:Gammas}
&&\Gamma_h =   -|A_h| \left(r U_{a}\Gamma_h +  \gamma U_b  \Gamma_e\right) \nonumber \\
&& \Gamma_e = -|A_h| \left(\gamma U_a \Gamma_e + U_b \Gamma_h \right).
\eea
In the main term we presented this equation and its solution (see below) for $r=1$.

The two eigenmodes of the set (\ref{eq:Gammas}) are
\be
\Gamma_h +\alpha_\pm \Gamma_e=\lambda^{++, +-}(\Gamma_h +\alpha_\pm \Gamma_e)
\ee
with
  \bea
  &&\alpha_{\pm} = -\frac{r-\gamma}{2}\frac{U_a}{U_b} \pm \sqrt{\left( \frac{r-\gamma}{2} \right)^2\frac{U_a^2}{U_b^2} + \gamma } \label{ch_2_1} \\
  &&\lambda^{++, +-} =- |A_h| \left[ \frac{ r+\gamma}{2}U_a \pm  U_b \sqrt{\left( \frac{r-\gamma}{2}\right)^2\frac{U_a^2}{U_b^2} + \gamma } \right], \nonumber
\eea

\section{Common part of the self-energies to order $\Gamma^2_{h,e}$}

We now show the details of the evaluation of the common part of the self-energies $\Sigma^Y_{xz}$ and $\Sigma_{yz}^X$ to second order in $\Gamma_{h,e}$.
   We can isolate the quadratic terms by evaluating $\Sigma_2=\Sigma_{xz}^Y+\Sigma_{yz}^X-2\Sigma^{tet}$.
    The common self-energy $\Sigma_2$ is given by
\be
\Sigma_2=U_M (n^Y_{xz} + n^X_{yz}) + U_{\Gamma} (n^\Gamma_{xz} + n^\Gamma_{yz})
\ee
where  $U_M = U_5 + 2 {\tilde U}_5 - {\tilde{\tilde U}_5}$ and $U_\Gamma = 2 (U_1 + {\bar U}_1) - (U_2 + {\bar U}_2)$.
The sums of the densities in the vicinity of the hole pockets are
\begin{align}
(n^\Gamma_{xz} + n^\Gamma_{yz})&= \int \frac{d^2k}{(2\pi)^2} \big( n_F(\epsilon_c^{nem}) +n_F(\epsilon_d^{nem})\big) + n^\Gamma_{tet} \notag \\
&= \Gamma_h^2\int \frac{d^2k}{(2\pi)^2}  \sum_{i=c,d} e^{\epsilon_i^{tet}/T}\left[\sigma_i\frac{1}{T b k_F^2}\frac{1}{\left(1+e^{\epsilon_i^{tet}/T}\right)^2} + \frac{1}{2T^2}\cos^2\theta \frac{e^{\epsilon_i^{tet}/T}-1}{\left(1+e^{\epsilon_i^{tet}/T}\right)^3}\right] + n^\Gamma_{tet}\notag\\
&=-\Gamma_h^2 \left[ \frac{bm_h^2}{1-b^2m_h^2} \frac{1}{2\pi b k_F^2} \left( \tanh{\left( 1+\frac{\epsilon_h}{2T}\right)} \right) + \frac{1}{8T \cosh^2\frac{\epsilon_h}{2T}}\right] + n^\Gamma_{tet}
\end{align}
where $\sigma_i=1 (-1)$ for $i=c(d)$,
 $\epsilon_h$ is defined after Eq. (\ref{ch_1_1}), and
  $n_{tet}^\Gamma$ is the density around hole pockets in the tetragonal phase.
 We define $(n^H_{xz} + n^H_{yz}) =  n^\Gamma_{tet} + B_h \Gamma^2_h$. Then
\be
B_h = - \left[ \frac{bm_h^2}{1-b^2m_h^2} \frac{1}{2\pi b k_F^2} \left( \tanh{\left( 1+\frac{\epsilon_h}{2T}\right)} \right) + \frac{1}{8T \cosh^2\frac{\epsilon_h}{2T}}\right]
\ee
We see that $B_h<0$.  We use this in the main text.
The densities around electron pockets are
\begin{align}
 (n^Y_{xz} + n^X_{yz})&=\int \frac{d^2k}{(2\pi)^2} \left[\cos^2\phi_Y n_F(\epsilon_Y) +\cos^2\phi_X n_F(\epsilon_X)\right] \nonumber \\
&= \int \frac{d^2k}{(2\pi)^2} \left[ A^2(1+\beta \Gamma_e)^2\sin^2\theta_Y n_F(\epsilon^{nem}_Y) + A^2(1-\beta \Gamma_e)^2\sin^2\theta_X n_F(\epsilon^{nem}_X)\right]\nonumber \\
&\approx A^2 \Gamma_e^2 \int \frac{d^2k}{(2\pi)^2} \sin^2\theta \left[ 2f_Y^{(2)}(k) n_F'(\epsilon_Y^{tet}) + \frac{1}{4}f_Y^2(k) n_F'' (\epsilon_Y^{tet}) +\beta f_Y(k) n_F'(\epsilon_Y^{tet}) \right] + (1+\beta^2\Gamma_e^2)n^{X,Y}_{tet}
\end{align}
In this  formula,  we used the symmetry between the expressions  for the densities at $X$ and $Y$ for $\Gamma_e=0$ and expanded to second order in $\Gamma_e$. We defined
$f_Y$ and $f_Y^{(2)}$ by writing $\epsilon_Y^{nem}=\epsilon_Y^{tet}+\Gamma_e/2 f_Y(k) +\Gamma_e^2 f_Y^{(2)}(k)$, where $f_Y$ is given below Eq.~\eqref{eq:ne}, and $f_Y^{(2)}=2v^2k_x^2/\sqrt{(C_1^Y-C_2^Y)^2+8v^2kx^2}$.
We then
define  $(n^Y_{xz} + n^X_{yz}) = n^{X,Y}_{tet} + B_e \Gamma^2_e$
 and  compute $B_e$ numerically using the  parameters for FeSe from  Ref.~\cite{kang_latest}).
 We find  $B_e<0$ and $|B_e|<|B_h|$.
   We use this in the main text.

As we said in the main text, this calculation is not fully self-consistent because we evaluated the "source" term by keeping the terms linear in $\Gamma_h$ and $\Gamma_e$ in the orbital Hamiltonian
 and expanding the densities to order $\Gamma^2_{h,e}$.
For a full self-consistent calculation, we should have
included also terms of order $\Gamma^2_{h,e}$ into the orbital Hamiltonian,
expanded to order $\Gamma_{e,h}^2$
and then solved self-consistently for the prefactor of the
$\Gamma^2_e$ term near the electron pockets.  However, the $\Gamma^2_e$ term
only appears because of the source term.
The prefactor is then proportional to the source term and has the same sign.  This is what we used in the main text.

\end{widetext}


\begin{thebibliography}{10}
\bibitem{footnote} In principle, one has to distinguish between interactions on hole and electron pockets for $U_a$. However, they differ only by a factor \cite{ruiqi}.
To smplify the presentation, in the main text we set $U_a$ to be equal on hole and  electron pockets. We discuss a generic case in~\cite{SM}.
\end{thebibliography}
\end{document}